\documentstyle[emulateapj]{article}
\newcommand{\lopt}{\ifmmode L_{2500} \else $~L_{2500}$\fi}
\newcommand{\loglopt}{\ifmmode{\rm log}~L_{2500} \else log$~L_{2500}$\fi}
\newcommand{\logz}{\ifmmode{\rm log}~z \else log$~z$\fi}
\newcommand{\ew}{\ifmmode{W_{\lambda}} \else $W_{\lambda}$\fi}
\newcommand{\ax}{\ifmmode{\alpha_x} \else $\alpha_x$\fi} 
\newcommand{\aox}{\ifmmode{\alpha_{ox}} \else $\alpha_{ox}$\fi} 
\newcommand{\kms}{\ifmmode~{\rm km~s}^{-1}\else ~km~s$^{-1}~$\fi}
\newcommand{\mone}{\ifmmode ^{-1}\else$^{-1}$\fi}
\newcommand{\mtwo}{\ifmmode ^{-2}\else$^{-2}$\fi}
\newcommand{\IUE}{{\em IUE}}
\newcommand{\HST}{{\em HST}}
\newcommand{\ciii}{\ifmmode{{\rm C\,III]}} \else C\,III]\fi}
\newcommand{\aliii}{\ifmmode{{\rm Al\,III}} \else Al\,III\fi}
\newcommand{\civ}{\ifmmode{{\rm C\,IV}} \else C\,IV\fi}
\newcommand{\heii}{\ifmmode{{\rm He\,II}} \else He\,II\fi}
\newcommand\lyb{\ifmmode {\rm Ly}\beta \else Ly$\beta$\fi}
\newcommand{\lya}{\ifmmode{{\rm Ly}\alpha}\else Ly$\alpha$\fi}
\newcommand{\Feuv}{\ifmmode{{\rm Fe_{UV}}} \else Fe$_{\rm UV}$\fi}
\newcommand{\mgii}{\ifmmode{{\rm Mg\,II}} \else Mg\,II\fi}
\newcommand{\ovi}{\ifmmode{{\rm O\,VI}} \else O\,VI\fi}
\newcommand{\siiv}{\ifmmode{{\rm Si\,IV}} \else Si\,IV\fi}
\newcommand{\Feo}{\ifmmode{{\rm Fe{\rm II}_{opt}}} \else FeII$_{\rm opt}$\fi}
\newcommand{\hb}{\ifmmode {\rm H}\beta \else H$\beta$\fi}
\newcommand{\oiii}{\ifmmode{\rm [O\,III]} \else [O\,III]\fi}
\newcommand{\woiii}{\ifmmode{{W_{\lambda}(\rm [O\,III])}} \else $W_{\lambda}$([O\,III])\fi}
\newcommand{\wciii}{\ifmmode{W_{\lambda}({\rm C\,III]})} \else $W_{\lambda}$(C\,III])\fi }
\newcommand{\wciv}{\ifmmode{W_{\lambda}({\rm C\,IV})} \else $W_{\lambda}$(C\,IV)\fi}
\newcommand{\wFeo}{\ifmmode{W_{\lambda}({\rm FeII_{\rm opt}})} \else $W_{\lambda}$(FeII_{\rm opt})\fi}
\newcommand{\wFeuv}{\ifmmode{W_{\lambda}({\rm FeII_{\rm UV}})} \else $W_{\lambda}$(Fe_{\rm UV})\fi}
\newcommand{\whb}{\ifmmode{ W_{\lambda}({\rm H}\beta )} \else $W_{\lambda}$(H$\beta$)\fi}
\newcommand{\wheii}{\ifmmode{W_{\lambda}({\rm He\,II})} \else $W_{\lambda}$(He\,II)\fi}
\newcommand{\wlya}{\ifmmode{W_{\lambda}({\rm Ly}\alpha)}\else $W_{\lambda}$(Ly$\alpha$)\fi}
\newcommand\wlyb{\ifmmode{ W_{\lambda}({\rm Ly}\beta )} \else $W_{\lambda}$(Ly$\beta$)\f
i}
\newcommand{\wmgii}{\ifmmode{W_{\lambda}({\rm Mg\,II})} \else $W_{\lambda}$(Mg\,II)\fi}
\newcommand{\wovi}{\ifmmode{W_{\lambda}({\rm O\,VI})} \else $W_{\lambda}$(O\,VI)\fi}
\newcommand{\wsiiv}{\ifmmode{W_{\lambda}({\rm Si\,IV})} \else $W_{\lambda}$(Si\,IV)\fi}
\newcommand{\lapprox }{{\lower0.8ex\hbox{$\buildrel <\over\sim$}}}
\newcommand{\gapprox }{{\lower0.8ex\hbox{$\buildrel >\over\sim$}}}

\received{January 26, 2001}
\accepted{}

\lefthead{Green, Forster, \& Kuraszkiewicz}
\righthead{Quasar Evolution and the Baldwin Effect}

\begin{document}
\title{Quasar Evolution and the Baldwin Effect in the \\
Large Bright Quasar Survey}

\author{Paul J. Green, Karl Forster, \& Joanna Kuraszkiewicz}
\affil{Harvard-Smithsonian Center for Astrophysics, 60 Garden St.,
Cambridge, MA 02138} 
\affil{email: {\em pgreen@cfa.harvard.edu, kforster@cfa.harvard.edu,
jkuraszkiewicz@cfa.harvard.edu}}

\begin{abstract}
From a large homogeneous sample of optical/UV emission line
measurements for 993 quasars from the Large Bright Quasar Survey
(LBQS), we study correlations between emission line equivalent width
and both restframe ultraviolet luminosity (i.e., the Baldwin Effect)
and redshift.  Our semi-automated spectral fitting accounts for
absorption lines, fits blended iron emission, and provides upper
limits to weak emission lines.  Use of a single large, well-defined
sample and consistent emission line measurements allows us to
sensitively detect many correlations, most of which have been
previously noted.  A new finding is a significant Baldwin
Effect in UV iron emission. Further analysis reveals that the primary
correlation of iron emission strength is probably with redshift,
implying an evolutionary rather than a luminosity effect.  We
show that for most emission lines with a significant Baldwin Effect,
and for some without, evolution dominates over luminosity effects.
This may reflect evolution in abundances, in cloud covering factors,
or overall cloud conditions such as density and ionization.  We find 
that in our sample, a putative correlation between Baldwin Effect
slope and the ionization potential is not significant.  Uniform
measurements of other large quasar samples will extend the luminosity
and redshift range of such spectral studies and provide even stronger
tests of spectral evolution.
\end{abstract}


\section{Introduction}
\label{intro}

  Because the simplest photoionization models for the emission-line
regions of quasars predict a linear proportionality between line and
continuum strength, diagnostics such as equivalent
widths (the ratio of integrated line flux over local continuum flux
density) would be expected naively to be {\em independent} of
continuum luminosity.  Baldwin (1977) first noticed that the
CIV~$\lambda1549$\AA\, emission line equivalent  
width ($W_{\lambda}$ hereafter) in quasars {\em decreases} with
increasing UV continuum ($1450$\AA) luminosity.  Since flux {\em
ratios} like \ew\, are distance-independent, this discovery brought
hope that, regardless of its physical explanation, the ensemble
Baldwin Effect (EBEff hereafter) might enable quasars (QSOs) to be
used as a ``standard candle'' in measuring cosmological distances. 

Unfortunately, the large dispersion in this anticorrelation 
(e.g., Baldwin, Wampler, \& Gaskell 1989; Zamorani et al. 1992) yields
poor distance calibrations relative to other standard candles. 
Relative luminosity distances accurate to 10\% at $z\sim0.5$ and 20\%
at $z=1$ are becoming possible with Type~Ia supernova measurements
(e.g., Garnavich et al. 1998; Perlmutter et al. 1999).  Combining
these results with those of Boomerang, COBE, Planck, and MAP should
more tightly constrain cosmological models in the near future (Park et
al. 1998; Melchiorri 2000).  A deeper understanding of the EBEff,
elusive though it still seems, is worth pursuing for several
reasons. First, QSOs represent the most distant non-transient
bright objects observable in the Universe, extending to $z\sim6$ (Fan et
al. 2000). Second, substantial reduction in the scatter is not 
critical to use of the EBEff for cosmology.  A `main sequence
fitting' approach with an ensemble of quasars could yield useful
cosmological constraints, as long as we can be confident that there
are no significant unaccounted for evolutionary effects (Baldwin
1999).  Third, the EBEff harbors some important information
on the nature and evolution of QSOs themselves.  If we can understand
both the origin of the EBEff and the source(s) of scatter, we
will have learned much about the intrinsic physics of QSOs.  

Physical processes or variables that could dominate the EBEff were
recently sketched by Sergeev et al. (1999) and Green (1999), and
include (1) Geometry - The inclination of an accretion disk could
change the apparent continuum luminosity alone (Wilkes et al. 1999; 
Netzer, Laor, \& Gondhalekar 1992).  (2) Covering factor - A decrease
in covering factor of broad emission line (BEL) clouds with luminosity
(Wu, Boggess, \& Gull 1983).  (3) Optically thin clouds - 
As the dominant ionization state of the element changes, a flat or
even {\em negative}  correlation between continuum flux and line
emission for a given species can result (Shields, Ferland \& 
Peterson 1995). (4) Changes in spectral energy distribution (SED) with
luminosity - A softer ionizing continuum in more luminous nuclei
(Green 1998; Korista et al. 1998; Wandel 1999a) causes a decrease in
emission line flux.  

Observationally, a significant EBEff has been claimed not only for
CIV, but also for ions such as OVI, He\,II, CIII], Mg\,II, and
Ly$\alpha$ (e.g., Tytler \& Fan 1992; Zamorani et al. 1992).  In all
cases, we refer to slopes in the log-log domain of equivalent width
$W_{\lambda}$ vs. luminosity $L$.  Thus, a slope $\beta_w$ describes
$W_{\lambda}\propto L^{\beta_w}$.  Typical slopes
\footnote{Some studies characterize the relationships as $W_{\lambda}\propto
M_V^{\beta_m}$ ($\beta_m=-0.4\beta_w$; Zamorani et al. 1992; Zheng,
Fang, \& Binette 1992), or $L_{line} \propto L^{\beta}$
($\beta=1+\beta_w$; Pogge \& Peterson 1992).}
for CIV are $\beta_w\sim -0.2$ (Kinney, Rivolo, \& Koratkar 1990) and
there have been claims that the EBEff shows steeper slopes for lines
of higher ionization energy (Zheng, Fang, \& Binette 1992; Espey \&
Andreadis 1999). In the SED picture, steeper slopes are naturally
expected for species of 
higher ionization energy, since shifts in the relative normalization
between their driving and underlying (UV) continuum are then more
effective.  An observational effect that has yet to be fully accounted
for is that the narrow component of emission lines varies most with
luminosity in samples of QSOs (Osmer, Porter, \& Green 1994).

A line/continuum anticorrelation analogous to the EBEff is also
observed in multi-epoch optical/UV spectroscopy of individual
active galactic nuclei (AGN) of lower luminosity ($M_B>-23$). The steep
slope of the intrinsic Baldwin effect (IBEff) ($\beta_w \sim -0.7$)
may add considerable scatter to the shallower 
EBEff, ($\beta_w \sim -0.2$; both slopes from Kinney, Rivolo, \&
Koratkar 1990), unless suitably time-averaged data are used for every 
object.   The scatter in the IBEff decreases once time lags
between emission line and continuum variations are removed (Pogge \&
Peterson 1992).  The IBEff appears to persist even into the
hard X-ray regime (for the Fe~K$\alpha$ emission line; Iwasawa \&
Taniguchi 1993; Reeves et al. 2001). While multi-epoch spectroscopy and
time-lag correction for large AGN samples might reduce scatter,
practically speaking, it is prohibitive.  Furthermore, variability
anticorrelates with luminosity in AGN (Helfand et al. 2001; Webb \&
Malkan 2000; Giveon et al. 1999), so that scatter from the IBEff is
unlikely to dominate in high luminosity QSOs. 

Unfortunately, many studies of the EBEff have been conducted using
compilations of measurements from the literature.  These compilations
mix samples that are diverse in their selection criteria, in the
resolution and quality of their spectra, and in the techniques used to
measure them.  Line measurement techniques for large samples suffer
from difficulty in achieving consistent and reliable
measurements of the continuum, and of blended line emission.  The
latter problem is particularly thorny for the blended iron multiplets
(FeI, FeII, and FeIII) and emission lines in close proximity.
\footnote{The optical iron emission appears to be predominantly FeII
(Boroson \& Green 1992).}
Absorption within the line profiles is rarely accounted for, and while
few studies provide upper limits to undetected emission lines, the
latter are  invaluable for confirming or constraining claimed trends.
We have therefore undertaken a major study of quasar line emission,
accounting for absorption lines and blended iron emission, using largely
automated procedures on carefully-selected samples, and providing
upper limits for undetected lines. The analysis of samples of QSOs with more 
than a few hundred spectra requires some amount of automatization to give
consistent results. The largest sample of QSO spectra currently
available is that of the Large Bright Quasar Survey.  Forster et
al. (2001; Paper\,I hereafter) describe the initial results from emission
line measurements of 993 LBQS QSOs (excluding those with strong broad
absorption). Here, we use these measurements to study the Baldwin
effect in the LBQS.  

\section{Significant Luminosity Correlations}
\label{beff}

Here we describe correlations between \ew\ and the restframe UV
luminosity.  We use \lopt, the monochromatic luminosity at 2500\AA\
obtained by extrapolating $B_J$ photometric magnitude and assuming a
continuum slope of $\alpha=0.5$ ($f_{\nu}\sim\nu^{\alpha}$).
Photometric magnitudes are superior to those 
estimated from the spectra themselves, since the LBQS spectra are not
spectrophotometric.  Use of luminosities calculated for other rest
wavelengths would simply offset the \loglopt\, value for each quasar
by a fixed amount (e.g., 26\% fainter at $\lambda1549$, 43\% fainter
at $\lambda 1216$), whereas we concern ourselves with EBEff slopes.
Optical luminosities are calculated assuming $H_0=50$ km s\mone
~Mpc\mone, $q_0=0.5$, and $\Lambda =0$, with further details in Green
et al. (1995).  

The \ew\, values and corresponding errors that we use are from
Paper\,I, but in cases where emission lines were modeled by more than
one Gaussian component (exclusively for strong lines like \lya, \civ,
and \mgii), we have combined those measurements, yielding a single
\ew\, measurement for all objects.  The median per-pixel
signal-to-noise ratio (S/N) of the LBQS spectra is typically $\sim$5,
independent of redshift.  Our emission line fitting procedure yields
uncertainty estimates, all of which are available in Paper\,I.  The
resulting emission line S/N can be characterized by ratio of the \ew\,
values to their $1-\sigma$ errors.  Representative median (mean) S/N
for strong UV line measurements are 5-6 (6-10).  Weaker lines (e.g.,
H$\gamma$, O\,I$\lambda1305$, O\,II$\lambda 3728$,
\siiv$+$OIV]$\lambda 1400$) have S/N between 3-5.  For emission lines
that are not detected, we include 2$\sigma$ upper limits to \ew\ using
the survival analysis package ASURV (Lavalley, Isobe, \& Feigelson
1992), and we apply the following tests to each pair of parameters:
the Cox proportional hazard model, the generalized Kendall rank and
the Spearman rank test. The probabilities of a correlation occurring
by chance in these tests are presented in Table~1 (column 4), where
the probability obtained from the Cox, Kendall, and Spearman rank
tests are listed in that order. The number of line measurements
studied and the number of upper limits included is also shown. We
considered a correlation significant only if the probability of a
correlation occurring by chance in all tests was less than 1\% and the
fraction of upper limits was less than 1/3. For comparison we also
calculate the probability of a chance correlation for data that
incorporate only detections. In Table~1 we also quote the slopes and
the intercept coefficients for correlations, calculated using both all
data and detections only.


We find significant correlations between \lopt\, and the equivalent
widths of the following emission lines: Ly$\alpha$, Si\,IV +
O\,IV]\,$\lambda$1400, C\,IV, Al\,III, Mg\,II.  These correlations are
displayed in Figure~\ref{fig-beff}, with their correlation and
regression results in Table~1.  The slope of the Ly$\alpha$ Baldwin
effect in our sample ($\beta_w{\rm (\lya)}=-0.27 \pm 0.07$) is steeper
than the slope of $-0.12 \pm 0.05$ obtained by Kinney et al. (1990),
for a compilation of 114 \IUE\, spectra combined with measurements
from several previously published studies.  Our measured slope is also
steeper than that obtained by Espey \& Andreadis (1999; EA99
hereafter), who analyzed the \IUE\, and \HST\, spectra of a
heterogeneous sample of 200 AGN to obtain $-0.08 \pm 0.03$. This may
be the result of a smaller luminosity range covered by our sample (1
dex) compared to the other samples (4-5 dex), since it has been noted
that the slope of the Baldwin effect flattens with increasing
luminosity range (e.g. Netzer, Laor, \& Gondhalekar 1992).  For the
C\,IV and Mg\,II Baldwin effect, our luminosity range is slightly
larger than that for \lya, and we still find slopes that are slightly
steeper than some previous studies.  By comparison, $\beta_w{\rm
(CIV)}$ was found to be $-0.17 \pm 0.04, -0.17\pm 0.03$, and $-0.13
\pm 0.03$ by Kinney et al. (1990), EA99, and Zamorani et al. (1992)
respectively, while we find $-0.23 \pm 0.02$. Slopes $\beta_w{\rm
(MgII)}$ were found to be $+0.01 \pm 0.04$ and $-0.08 \pm 0.03$ by
EA99, and Zamorani et al. (1992) respectively, while we find $-0.19
\pm 0.02$.

We note that sensitivity to detecting any significant correlation is
affected not only by the S/N of the spectra, but also by the
luminosity range spanned by each emission line in the sample.  The
spectral coverage and flux limits of any uniform sample will yield a
maximum redshift/luminosity range for lines whose rest wavelengths are near
the short wavelength end of the observed-frame spectra. Consequently,
for the LBQS spectra, the luminosity range of Mg\,II and \Feuv\, are
best (2.6dex), while shorter wavelength lines like \lya\, span
$\sim1$dex, and optical lines (e.g., \hb, \oiii) span $\sim1.5$dex in
luminosity.  Several studies cited above present larger luminosity
ranges for some of the lines we investigate here.  In the present
work, we confine ourselves primarily to homogeneous measurements of a
uniformly-selected sample.  A followup paper will combine 
similar measurements of other samples to cover larger portions
of the $L-z$ plane.

Most studies have shown little evidence for an EBEff in the SiIV+OIV]
blend, although here we confirm with $\sim$400 QSOs the significant
correlation noted by Laor et al. (1995) from \HST\, spectroscopy of
just 14 QSOs.  We do not confirm the Baldwin effect found by Zheng,
Kriss \& Davidsen (1995) in Ly$\beta +$\ovi\, measurements. However,
if we analyze only the detections, we do find a significant
correlation (see Table~1 and Figure~{\ref{fig-beffo6}}a). The slope of
this correlation ($-0.50 \pm 0.19$) is consistent within the errors
with the slope found by Zheng, Kriss \& Davidsen (1995; $-0.30 \pm
0.03$). These results show that including upper limits in the analysis
can help to avoid spuriously significant correlations. However, as
noted by Green (1998) the Zheng et al. result is probably still valid,
as the O\,VI emission was detected for every QSO in their sample and
no upper limits have been ignored in the analysis.  This may imply that
we find no O\,VI Baldwin effect due to a smaller luminosity range
covered by our sample (1 dex cf. 5 dex in Zheng et al.). 

We also find a marginal Baldwin effect for H$\delta$ and 
He\,II$\lambda 4686$, where 72\% and 83\% of the data respectively are 
non-detections. It is, however, reassuring that Wilkes et al. (1999),
who study a heterogeneous sample of X-ray bright AGN, find the
H$\delta$ Baldwin effect for a sample which includes only 30\% 
non-detections.  We also find a marginal Baldwin effect for the
following forbidden lines [Ne\,V], [O\,II], [Ne\,III] and a trend 
for the [O\,III] equivalent width to correlate with \lopt.  
We call these trends marginal because the measurements consist of
50-75\% upper limits, but they are worthy of further investigation
in samples of higher S/N.

We do not confirm the CIII] Baldwin effect first reported as an IBEff by
Zheng, Fang \& Binette (1992) for Fairall~9 and then by Green (1996)
as a EBEff for a sample of 85 QSOs with \IUE\, spectra.  While the
CIII] EBEff has $P_{Cox}<0.01$, it does not meet our criteria
for the other correlation tests. We also cannot
confirm the He\,II\,$\lambda$1640 Baldwin effect found by Green
(1996).  Neither do we see a significant EBEff for NV, similar
to Korista et al. (1998).  A shallow, or even {\em positive
correlation} with luminosity has been reported elsewhere for NV
(EA99), and has been attributed to abundance effects (Hamann \&
Ferland 1993, HF93 hereafter).  We discuss this further below in
\S~\ref{evol} and \S~\ref{ioniz}. Unfortunately, in many AGN spectra,
NV is difficult to deblend from \lya (see also discussion in HF93).
The effects of blending are much less pronounced in \lya\ itself,
since it is by a factor of 3-4 the stronger of the two lines. 

\section{The Iron Baldwin Effect}
\label{ironbeff}

Many spectra show optical and ultraviolet (UV) iron emission blends
(e.g. Wills, Netzer, \& Wills 1985; Boroson \& Green 1992) which appear
most strongly around the H$\beta$+[OIII] complex and the
Mg\,II$\lambda$2798 line. We model this emission in our spectra by
independently fitting the optical template of Boroson \& Green (1992),
covering 4400\AA$ < \lambda_{rest} <$7000\AA, and and the UV template
of Vestergaard \& Wilkes (2001), covering 1250\AA$ < \lambda_{rest} <
3100$\AA.  For further details on the iron modeling see Paper\,I. We
report for the first time a significant Baldwin effect 
for UV iron emission ($P<10^{-4}$; 31\% of 953 measurements are upper
limits). We may be particularly sensitive to detecting the trend
because the range of luminosity over which we can measure \Feuv\,
emission in the LBQS sample is larger than for any other emission line
(about 3 orders of magnitude).  We note that most of the upper limits
occur for $z>1.4$, which corresponds closely to $\loglopt \gapprox 31$ in
the LBQS (see Figure~\ref{fig-lz}).  To confirm our results, we
excluded these objects from the sample and reran the ASURV analysis.
The BEff is still highly significant ($P<10^{-4}$; 16\% of 611
measurements are upper limits).  The correlation also remains strong
when only the (659) detections are analyzed. 

We find only a marginal Baldwin effect for optical Fe\,II emission
($P=10^{-4}$ but 44\% of the measurements are upper limits). The
optical iron correlation is not significant in the LBQS if only
detections are analyzed. Our current analysis emphasizes consistent
measurements of homogeneous samples, but we may also benefit by
combining with other such samples to extend the luminosity range. We
therefore analyze the only other such sample currently available to
us, the optical spectra of 87 QSOs - the Palomar-Green (PG) sample of
Boroson \& Green (1992).  The luminosity range is thereby just
slightly extended down to $\loglopt =28.94$.
We measured the strengths of H$\beta$, [O\,III], He\,II$\lambda 4686,$
and \Feo\, emission using the automated technique described in
Paper\,I.  The combined PG and LBQS sample detects a
significant Baldwin effect for \Feo\, (where 33\% of data
in the combined sample are comprised of upper limits).  
No other new significant trends are detected in the PG+LBQS sample
according to our criteria.\footnote{The He\,II$\lambda 4686$
EBEff is significant, but the upper limits for He\,II decline from
83\% in the LBQS to 58\% in the combined sample, still above our
threshold of 1/3.}

\section{Evolution of Iron Emission Strength}
\label{ironevol}

As expected in a flux-limited sample such as the LBQS, luminosity and
redshift are strongly correlated, and the available range of \lopt\,
decreases with redshift.   A plot of \loglopt\, vs. log-redshift is
shown in Figure~\ref{fig-lz}.  In the LBQS, \lopt\, spans a factor of
$\sim 50$ for $z<0.5$, 20 up until $z\sim2$, and $\sim8$ for $z>2$.
Does iron emission strength depend more on luminosity, or redshift?
This question has not been discussed much in treatments of the EBEff
in general (though see Baldwin, Wampler, \& Gaskell 1989).  This is 
surprising, since the finding
of a stronger redshift correlation would carry the important
implication that evolution is directly detectable in quasar spectra.
Here we take advantage of our large sample and consistent measurements
to test for the primary relationship. Astronomical measurements often
correlate strongly with several variables, and multivariate
statistical analyses have been performed in a number of studies in the
literature.  Examples include Green (1996), Eskridge, Fabbiano, \& Kim
(1995), Wilkes et al. (1994), and Djorgovski et al. (1993).
Partial Spearman Rank Analysis (hereafter PSRA; Kendall \& Stuart
1976) allows for correlation analysis in the general multivariate
case, using a matrix of  bivariate Spearman rank statistics as input.
PSRA tests for correlations between subsamples of the matrix
parameters while holding constant all other variables in the matrix.
We use the ASURV bivariate Spearman Ranks as input to multivariate
PSRA, and find that since the partial correlation coefficients
with redshift are of larger magnitude\footnote{Or correspondingly, the
probability $P$ of a null correlation, is significantly smaller when
redshift rather than luminosity is allowed to vary.}
{\em iron equivalent width in the UV is primarily anticorrelated with redshift, not with luminosity}. $\wFeuv$
anticorrelates most strongly with redshift ($P_{PSR}<0.005$, 
and a PSR of $\rho=-0.123$, while its correlation with \lopt\, has
$P_{PSR}=0.177$ and $\rho=-0.033$).  To check if this result is
plausibly related to the magnitude limit of the LBQS ($B_J=18.85$), we
also tested a brighter ($B_J<18.6$) subsample of the LBQS, and found a
similar result (see Table~2).  We discuss this subsample further
in \S~\ref{evol} below.

Our finding of iron evolution is supported by the recent comparison by
Kuhn et al. (2001) of two QSO samples {\em matched in evolved luminosity}.
\footnote{Both their high and low redshift samples span
the range $1< L/L_*(z) < 7$, where $L_*(z)=L_*(1+z)^k$,
the optical luminosity function parameterization of pure luminosity
evolution from Boyle, Shanks, \& Peterson (1988).}
They also found evidence for a decrease with redshift of the strength 
of the 2200-3000\AA\, bump.  

While iron is produced in all supernovae (SN), the dominant producer
appears to be SN~Ia (HF93; Wheeler et al. 1989).  Since SN~Ia
progenitors have a lifetime of $\sim$1~Gyr, large increases in iron
abundance could be delayed for at least that amount of time after the
first epoch of star formation. For $q_0=0.5$, the age of the Universe
is 1($h\mone$)Gyr at $z\sim3$. Thus as proposed by HF93, detection of
this abundance shift might enable cosmological tests.  In contrast to
iron, magnesium production should be dominated by type II, Ia, and Ib
supernova, so that changes in the Fe/MgII ratio should be most 
evident around the epoch when SN~Ia become prevalent.  Thompson, Hill,
\& Elston (1999) designed a test for this effect comparing three
composite spectra: two averages of 6 QSOs each with mean redshifts of
$\overline z\sim 3.4$ and $\overline z\sim 4.5$, and an LBQS composite
with $\overline z\sim 0.8$.  They found that the Fe/MgII was constant
within their measurement errors.  However, this ratio is certainly
complicated by the strong MgII EBEff that we and others detect.  When
Thompson et al. attempted an independent iron measurement, they found
$\wFeuv$ to be ``marginally larger'' in the highest redshift composite.
Unfortunately, the detailed structure of composite spectra depends
on how and where their constituent spectra are normalized
before adding (Brotherton et al. 2001), and this is especially
crucial when measuring blended iron emission around the strong, broad
MgII line. 

\section{Evolution of Emission Line Strength}
\label{evol}

Given the intriguing results showing that evolution of iron emission
strength is a primary correlation (with luminosity secondary) we
have pursued similar partial correlation analyses for all lines 
in our sample.  Partial correlation results are presented in Table~2
for all lines showing a significant correlation with
either luminosity or redshift.  For the full LBQS sample, the
influence of evolution on emission line strength appears to dominate
for all lines, since the probability $P$ of a null correlation is
significantly smaller when redshift rather than luminosity is allowed
to vary.  The equivalent widths of \lya, \civ, \aliii, \siiv,
\ciii, \mgii, \Feuv, and \hb\, all decrease with increasing redshift,
with a stronger dependence on redshift than on luminosity.  
The evolution regressions for all significant 
\hbox{\ew-redshift} correlations in the full sample are listed 
in Table~3, and drawn over the data (as dashed lines)
in Figure~\ref{fig-evol} for correlations that show
significant correlations separately with both luminosity and redshift.
Identical ranges in log\ew\, are used as in Figure~\ref{fig-beff} to
allow a direct comparison.  We perform a simple analysis that
we find to be an intuitive reflection of the visual impression
that evolution plots have less dispersion than EBEff plots.
We subtract the best-fit regression for those correlations also with a
significant EBEff in Table~1, and derive the residual RMS
dispersion.\footnote{Since the RMS is difficult to define when
including upper limits, we examine detections only, and use the
relevant regressions listed in Table~1 and Table~3.}  We find smaller 
RMS dispersions generally for the correlation we designate
as primary from the PSRA results in Table~2.  For this exercise, we
only compared correlations that were significant both for the EBEff and for
evolution.  Neither \ciii\, nor \hb\, are included, since we
detected no significant trend with luminosity.  Their correlations
with redshift are significant, and are plotted in Figure~\ref{fig-evolb}. 

The distribution of \ew\, upper limits is typically weighted towards
high luminosities and high redshifts, which if it is an observational
effect could bias a correlation or a regression when the fraction of
limits is large.  To test the effect of this potential bias, we also
ran partial correlation analyses on \ew\, as a function of \loglopt\,
and \logz\, for the subset of data that excludes line upper limits.
The correlation results for this sample are also shown in Table~2.
The primary correlation appears to still be redshift for  \lya, \siiv,
and \civ. The \Feuv\, results for detections only are ambiguous, since
the correlation coefficients ($\rho$) for \logz\, and \loglopt\, are
identical.  In any case, since exposure times for the LBQS spectra
achieved very similar median S/N ratios across the range of $B_J$ mag,
the appearance of weaker lines at higher redshift is most likely due
to an intrinsic effect. 


Similarly, since our results are plausibly biased by the magnitude
limit of the LBQS ($B_J=18.85$), we also tested a brighter
($B_J<18.6$) subsample of the LBQS.  Given that the magnitude range of
this bright subsample is from 16 to 18.6, it only covers about an
order of magnitude in blue flux, and therefore samples a different
part of the luminosity function at each redshift. However, the effects
of Malmquist biases and pileup at the sample flux limit are relieved,
since every true $B_J=18.6$ source was easily detected.  The mag
errors are also lower than at the survey limit, so that fewer faint quasars
land in the brighter subsample due only to photometric errors,
and there is less random error in the derived \loglopt\, values.  For
the $B_J<18.6$ subsample, all correlations which show an EBEff have
their primary correlation with redshift (Table~2).\footnote{However,
\wciii\, correlates significantly with neither luminosity ($P_S=0.081$)
or redshift ($P_S=0.045$) in the $B_J<18.6$ subsample.}  

\section{Trends with Ionization Potential}
\label{ioniz}

Zheng, Fang \& Binette (1992) analyzed the \IUE\, spectra of the
highly variable AGN Fairall~9 and found that the slope of the
intrinsic Baldwin effect becomes steeper for lines with higher
ionization potential.  Zheng, Kriss \& Davidsen (1995) found a similar
trend in a small sample of 32 QSOs and Seyfert~1 galaxies (observed by
\IUE, \HST, and HUT).  The EBEff showed the steepest
slope for the O\,VI line (with the highest potential),
followed by the slopes of C\,IV and Ly$\alpha$. This finding was later
confirmed by EA99, who studied a larger number of UV and optical line
slopes for a heterogeneous sample of $\sim$200 AGN
and found a trend ($P < 5\%$) that lines with increasing ionization
potential show a steeper Baldwin effect slope (see their
Figure~4). 

We present the dependence of the EBEff slope on ionization potential
in our LBQS sample in Figure~\ref{fig-espey}. Open circles represent
the slopes calculated for all LBQS data (from Table~1) with our linear
fit shown by a dashed line. The slopes we calculate for detections
only are represented by filled squares, with the best-fit slope shown
with a solid line. Our best-fit linear ``slope-of-slopes'' (SOS)
regressions for detections only ($-0.0026\pm0.0007$) and all data
($0.0003\pm0.0237$) are consistent within the errors with that found by
EA99 ($-0.0015\pm0.0005$).  However the correlation between slope and
the ionization potential is not significant in our sample, with or
without emission line upper limits included ($P=90\%$ and $P=30\%$
respectively).  The lack of a correlation is in part the result of
relatively steep EBEff slopes we measure for the low ionization
potential lines Ly$\alpha$ and \Feuv.  The steep Ly$\alpha$ slope we
measure is probably caused by the rather small luminosity range
covered for this wavelength in the LBQS sample (see Netzer et
al. 1992).  No \Feuv\, slope was included in EA99.  As described in
\S~\ref{ironbeff}, an \Feo\, EBEff is excluded as insignificant by our
adopted criteria, unless the PG sample is included. Inclusion of the
PG sample shifts the \Feo\, EBEff slope from $-0.45$ (see Table~1) to
$-0.28$, but the resulting SOS relation remains insignificant ($P \sim
90\%$).

  It would be intriguing if ionization potential yielded a
significant correlation, since the production of many emission lines
may be sensitive to continuum photons both softer and harder than the
ionization potential of the species in question, because such photons
may ionize from excited states and also heat the gas via free-free and
H$^-$ absorption.  For example, in most photoionization models (e.g.,
Krolik \& Kallman 1988), UV Fe originates at low optical depths in
BLR clouds where EUV and soft X-ray photons contribute to heating.
FeII emission in the optical is principally due to higher energy
photons (above $\sim 800$eV).  Ionization from
excited states and heating via free-free and H$^-$ absorption also
help determine the principal ionizing/heating continuum
(Krolik \& Kallman 1988) of an emission line.  Perhaps more
representative energies than the ionization potential should be used
in the SOS plot, or lines that respond most significantly to heating
could be excluded. The former involves many changes and considerable
uncertainty in the chosen energies, while the latter does little to
forge a significant correlation.  

We do not include a NV EBEff slope in the above discussion, because it
was not significant in the LBQS.  The N\,V line was excluded from the
slope of slopes figure of EA99 as well, because it did not fit
the overall relation. It was claimed that the aberrant slope  of NV is
due to abundance changes with redshift (Korista et al. 1998; Hamann \&
Ferland 1999), which we discuss briefly below.   

\section{Discussion}
\label{discussion}

The observed decrease of \ew(\Feuv) with redshift, if confirmed,
corresponds to an increase of $\wFeuv$ with cosmic time that could be
attributable to a number of evolutionary effects including: (1) a
increase in iron abundance (2) an increase in the covering factor of
iron-emitting clouds or (3) a shift in continuum SEDs or (4) of gas
conditions (i.e. density, temperature, or ionization) in the emitting
clouds.  If the significant evolutionary trends that we detect in this
and other measured emission lines are the result of abundance changes,
they proceed with the arrow of cosmic time in the expected sense;
abundances increase with time as more stars cycle metals into the
interstellar medium, thereby enhancing the abundances of clouds in the
nuclear environment of quasars.  However, most previous claims of
detected trends in the metallicity of QSO emission line clouds appear
to go in the opposite sense (Hamann \& Ferland 1999; Vernet et
al. 2001).  Those less intuitive trends, if true, might be explained
if QSOs at higher redshifts are more luminous (generally true in
flux-limited samples) and also more massive, analogous to the
mass-metallicity trend observed in nearby elliptical galaxies (e.g.,
K\"oppen \& Arimoto 1990).  Another explanation might be that the most
effective metallicity enhancements occurred at early cosmological
epochs, but high redshift QSOs are short-lived and unrelated to
low-redshift counterparts (HF93).  Are abundances larger in the BLR at
early or late epochs?   A determination of the correct answer is
important, since it provides a potential measure of the epoch and
lifetime of nuclear activity in galaxies.  

Line ratios may be more robust indicators of metallicity then
the equivalent width trends we emphasize within the scope of this
paper.  The measurements we provide in Paper\,I provide the basis for
our pursuit of further line ratio studies.  Nitrogen intensity should
be particularly sensitive to metallicity since as a secondary element,
it goes up roughly as the square of the metallicity ($Z$) in scenarios
of rapid star formation.  Expected line ratios for solar metallicity
gas are $\sim0.1$ for NV/CIV and $\sim1$ for NV/HeII (Hamann \&
Ferland 1999).  Ferland et al. (1996) found some robustly large
NV/HeII ratios in luminous QSOs, requiring $Z\gapprox 5Z_{\odot}$.
Dietrich \& Wilhelm-Erkens (2000) similarly derive $Z\gapprox
8Z_{\odot}$ for a sample of 16 QSOs ($2.4<z<3.8$).The correlation of
NV/CIV with NV/HeII (Hamann \& Ferland 1999; Vernet et al. 2001) seems
likely to be an abundance effect (Villar-Martin et al. 1999).   

There are caveats, however. NV and CIV are produced in different
regions, and their ratio is only linear in $Z$.  Because of the
substantial cooling afforded by CIV, an increase in abundance could be
accompanied by a decrease in temperature, which in turn reduces the
response of line strength to abundance; weak lines respond better.
While NV and HeII lines arise in regions of similar ionization, both
are notoriously difficult to measure in the majority of QSOs because
they are broad and often strongly blended with nearby lines.
Perhaps more importantly, we reiterate that the measurements of NV that 
dominate discussions of abundance are generally very difficult due to
blending with the much stronger Ly$\alpha$ line.  The proximity of NV
to Ly$\alpha$ also means that outflowing clouds (e.g., as seen in
BALs) may boost NV emission because of resonant scattering of
Ly$\alpha$ in the restframe of the cloud (Krolik \& Voit 1998).
Studies of broad absorption lines in quasars may help (Korista et
al. 1996).  These also suggest high BAL cloud abundances of up to
$10Z_{\odot}$, where rapid star formation models yield better 
abundance fits than do scaled solar metallicities.  The interpretation
of BAL measurements is in flux, however, since the BAL profiles seem
to be determined more strongly by partial covering than by optical
depth (Arav et al. 1999). 

Is the detected trend of iron an abundance effect?  Within
photoionization models,  the effect of abundance on iron line
strength is very weak due to the thermostatic effect of Fe\,II.  
Indeed, photoionization models have severe trouble accounting for the
observed strength of iron emission (Collin \& Joly 2000), requiring
the iron emitting region to be heated by an additional, non-radiative
mechanism.  Wind models look promising because outflows (1) could
produce the shocks and consequent non-radiative heating; (2)
may shield the narrow line region (NLR) or even replace it with a
denser medium.  Furthermore, there are analogies in stellar winds 
that are observed to produce both a sort of intrinsic
Baldwin Effect (Morris et al. 1993) and strong iron emission (e.g.,
Hillier \& Miller 1998).  So the underlying cause of the observed line
evolution might be evolution in outflows (e.g., Murray \& Chiang
1998), which may in turn be caused by evolution of black hole mass and
accretion rate (e.g., Wandel 1999b).  Studies of broad line width as a
function of look-back time can help address this question.  

Because evolution predominates in our sample even for \wlya, the
trends we detect are indeed more likely to correspond to evolution in
cloud conditions rather than in abundance (although the two are linked by
thermostatic effects).  Each QSO observable is an axis in a
multidimensional space where  that can be transformed 
using principal component analysis (PCA) into a new basis
space whose first eigenvector represents most of the diversity in QSO
spectra, and is a linear combination of the original axes
(observables) .  About 50\%\, of QSO optical/UV  spectral diversity
can thus be projected along a principal eigenvector of spectral
properties dubbed Eigenvector~1 (or PC1; Boroson \& Green 
1992; Wills et al. 1999, and references therein).  Linking such
measurables as FWHM(\hb), FeII/\hb, SiIII]/CIII], \wciv, \woiii, and
X-ray spectral slope \ax, PC1 has been hypothesized to reflect
accretion rate $L/L_{Edd}$ and/or the presence of outflowing winds.
Narrow line Seyfert~1s (NLS1s) and low-ionization broad absorption
line (loBAL) quasars share several properties that appear to lie at
one extreme of PC1.  Among them are weak narrow line but strong iron
emission, narrow FWHM(\hb), and perhaps evidence of outflows (Brandt
2000; Mathur 2000a) and steep (soft) intrinsic \ax (Mathur et
al. 2001).  One possibile explanation of larger \Feuv\, equivalent
widths in the present epoch is that outflows are now more 
common. Debate has begun on whether the accretion rate is large in the
early or late phases of evolution (Mathur 2000b; Wandel 1999c; Wilman
\& Fabian 1999).  If outflows indeed dominate \Feuv\, line
strength, then our work here suggests the latter.

We do not propose that redshift/emission line
correlations reported here completely explain the ensemble Baldwin
effect.  Significant EBEffs have been seen in samples spanning very
small redshift ranges (e.g., Netzer et al. 1992; the $0.08<z<0.4$
sample of Wills et al. 1999).  Furthermore, as described above
(\S~\ref{intro}, even individual AGN observed at multiple epochs (a
`sample' with zero redshift range) exhibit an intrinsic
\ew\,-luminosity anticorrelation (Kinney et al. 1990; Pogge \&
Peterson 1992).  Rather, here we find new evidence that evolution may
also play a role in the EBEff, and could be the primary correlation
for several important lines.  The simplest use of the EBEff to
constrain cosmological parameters like $q_0$ is similar to main
sequence fitting of star clusters, in that it depends on the
assumption of no (or at least predictable) evolution (Baldwin 1999).
Since evolution of the quasar luminosity function is well-accepted, it
is reasonable to also expect evolution in other observables like
emission line strength.  Rather than using quasars to measure $q_0$,
it will likely turn out instead that other methods of measuring
cosmological parameters (see \S~\ref{intro}) will provide a cosmology
sufficiently precise that we can then make more rapid progress
understanding quasar evolution, particularly at high redshifts.

The current study suffers from several problems.  First, the typical 
S/N of the LBQS spectra are too low, with a median of $\sim5$
averaged over the entire observed spectrum. Second, the
sample has a single relatively bright flux limit, so that $L$ and $z$
are strongly correlated. The ideal study would measure two large
quasar samples at very different redshifts, each spanning a wide but
similar range of $L$,  selected without regard to emission line
strength, and with high S/N spectra available.  As long as we're
making a wish list, multi-epoch spectroscopy for such samples would
also be valuable, for reasons outlined in the introduction.
Variability anticorrelates with luminosity in AGN (Webb \& Malkan
2000; Giveon et al. 1999), and there is no evidence that
variability correlates in any way with redshift (Helfand 2001;
Hawkins 2000).  

Extension of this study of the LBQS sample to lower redshifts and 
luminosities using consistent measurement techniques
is clearly of interest to fill in the luminosity-redshift
plane of Figure~\ref{fig-lz} and  alleviate the degeneracy
imposed by the strongest correlation in the current sample.
We are pursuing such a study by measuring a large but heterogeneous
sample of \HST\, FOS spectra (Kuraszkiewicz et al. 2001), and
expect that the results should prove convincing since the spectral 
coverage includes most of the same UV emission lines analyzed here.
The emission line properties of the optically-selected quasars from the Sloan
Digital Sky Survey (SDSS - Richards et al. 2001; York et 
al. 2000) could provide a large sample of great diversity.  
On the other hand, optical samples may highlight only  the brief
juncture in QSO evolution when luminosities are still large, but
enshrouding material has been blown away (Wilman \& Fabian 1999). 
Spectroscopy of upcoming X-ray selected samples from
Chandra (ChaMP; Green et al. 1999; Wilkes et al. 2001), similarly
analyzed, holds great promise to unravel the complexities of quasars'
intrinsic physics from their evolution over the span of the observable
Universe.   

The authors gratefully acknowledge support provided by NASA through
grant NAG5-6410 (LTSA).  We are grateful to Craig Foltz for providing
the LBQS spectra, to Todd Boroson for providing the spectra of the
PG QSOs, and to Jack Baldwin for providing his comments
on a draft of this manuscript.

\clearpage
\begin{figure}[ht]
\plotfiddle{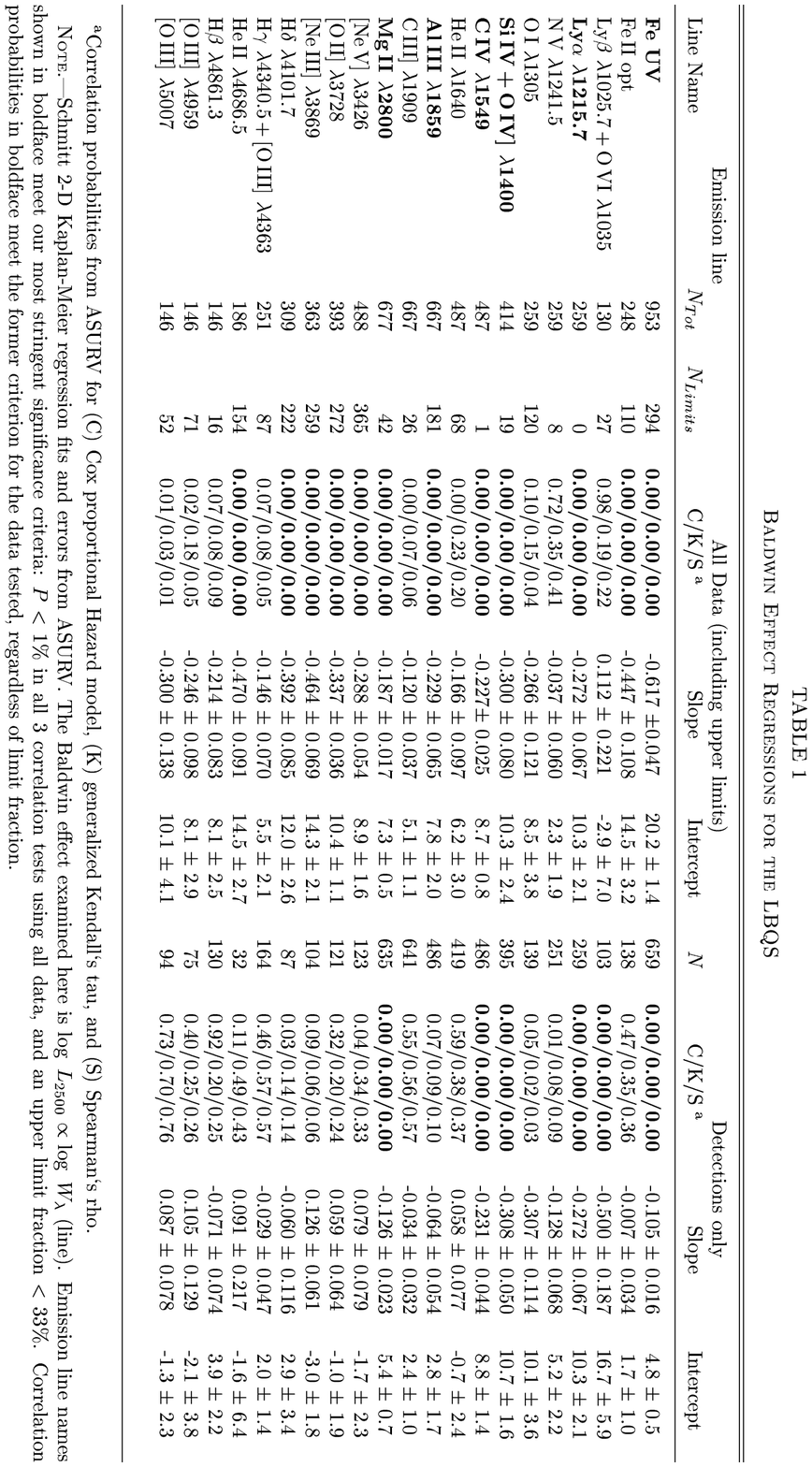}{14cm}{180}{100}{100}{350}{420}
\end{figure}

\voffset=-3cm
\begin{deluxetable}{lrcrccrcrc}
\tablewidth{0pt}
\tablenum{2}
\tablecaption{Partial correlations}
\tablehead{
\multicolumn{1}{c}{}&
\multicolumn{4}{c}{All data (including upper limits)}&
\multicolumn{1}{c}{\ \ }&
\multicolumn{4}{c}{Detections only} \\
\multicolumn{10}{c}{} \\
\colhead{Line}&
\multicolumn{2}{c}{log\ew\, vs. log L$_{2500}$}&
\multicolumn{2}{c}{log\ew\, vs. log z} &
\multicolumn{1}{c}{  }&
\multicolumn{2}{c}{log\ew\, vs. log L$_{2500}$}&
\multicolumn{2}{c}{log\ew\, vs. log z} \\
& $P~~$ & $\rho$ & $P~~$ & $\rho$ & & $P~~$ & $\rho$ & $P~~$ & $\rho$ 
}
\startdata
&&&&&&&&&\\
\multicolumn{10}{c}{LBQS Sample} \\
Fe UV	& 0.177 & -0.033&   {\bf $<$0.005} & -0.123 & & 0.175
	      & -0.983&    0.175 & -0.983\\
Ly$\beta +$O\,VI    &...&... &...&...&&{\bf  0.258} & -0.065&    0.358 & -0.037\\
Ly$\alpha$	& $<$0.005 &  0.210&  {\bf $<$0.005} & -0.301& &$<$0.005  & 0.207&  {\bf $<$0.005} & -0.298\\
Si\,IV + OIV]   &$<$0.005 & 0.280&  {\bf $<$0.005} & -0.382& &$<$0.005  & 0.300&  {\bf $<$0.005} & -0.409\\ 
C\,IV	& 0.250 & -0.031& {\bf  0.158 }& -0.048&&  0.245  & -0.032& {\bf  0.153} & -0.049\\ 
Al\,III	&$<$0.005 & 0.142&  {\bf $<$0.005 }& -0.207&&... &...&...&...\\
C\,III]    & 0.066 & 0.060&  {\bf  0.011 }& -0.091&&...&...&...&...\\ 
Mg\,II	& 0.154 &  -0.045& {\bf 0.075} & -0.063&&{\bf 0.011}&-0.092&0.305&0.021\\
H$\beta$&$<$0.005&0.313&{\bf $<$0.005 }&-0.378& &$<$0.005 & 0.286& {\bf $<$0.005 }&-0.338\\
&&&&&&&&&\\
\multicolumn{10}{c}{LBQS $B_J<18.6$ Subsample}\\
Fe UV	& 0.177 & -0.033& {\bf $<$0.005} & -0.123&&
 0.244  & -0.030&   {\bf  0.208} & -0.037\\
Ly$\alpha$     &$<$0.005 & 0.379&  {\bf $<$0.005} & -0.446 &&$<$0.005 & 0.161&  {\bf $<$0.005} & -0.258\\
Si\,IV + OIV]   &...&...&...&...&&$<$0.005 & 0.164&  {\bf $<$0.005} & -0.272\\
C\,IV     & 0.239 & 0.037&  {\bf  0.024} & -0.103&& 0.244  & 0.036&  {\bf  0.023} & -0.104\\  
Al\,III	&$<$0.005 & 0.151&  {\bf $<$0.005} & -0.204&&...&...&...&...\\
Mg\,II    & $>$0.400 & 0.000&  {\bf  0.012} &
-0.098&&{\bf  0.181}  & -0.041 &    0.257 & -0.028\\
H$\beta$&$<$0.005&0.520&{\bf $<$0.005} &-0.565&&$<$0.005&0.535& {\bf $<$0.005} &-0.568\\
\tablecomments{$P$ is the partial Spearman rank probability
and $\rho$ is the partial correlation coefficient
of that \ew\, correlation occurring by chance, given 
that \ew\, may depend on both log~$L_{2500}$
and log~$z$, and holding each of these variables constant
in turn.  Bold fonts denote the primary correlation for each line. No
values are listed for correlations that are not significant (i.e.,
have $P<0.01$ and $<33\%$ upper limits) in  either independent
variable singly.} 
\enddata
\end{deluxetable}

\begin{figure}[ht]
\plotfiddle{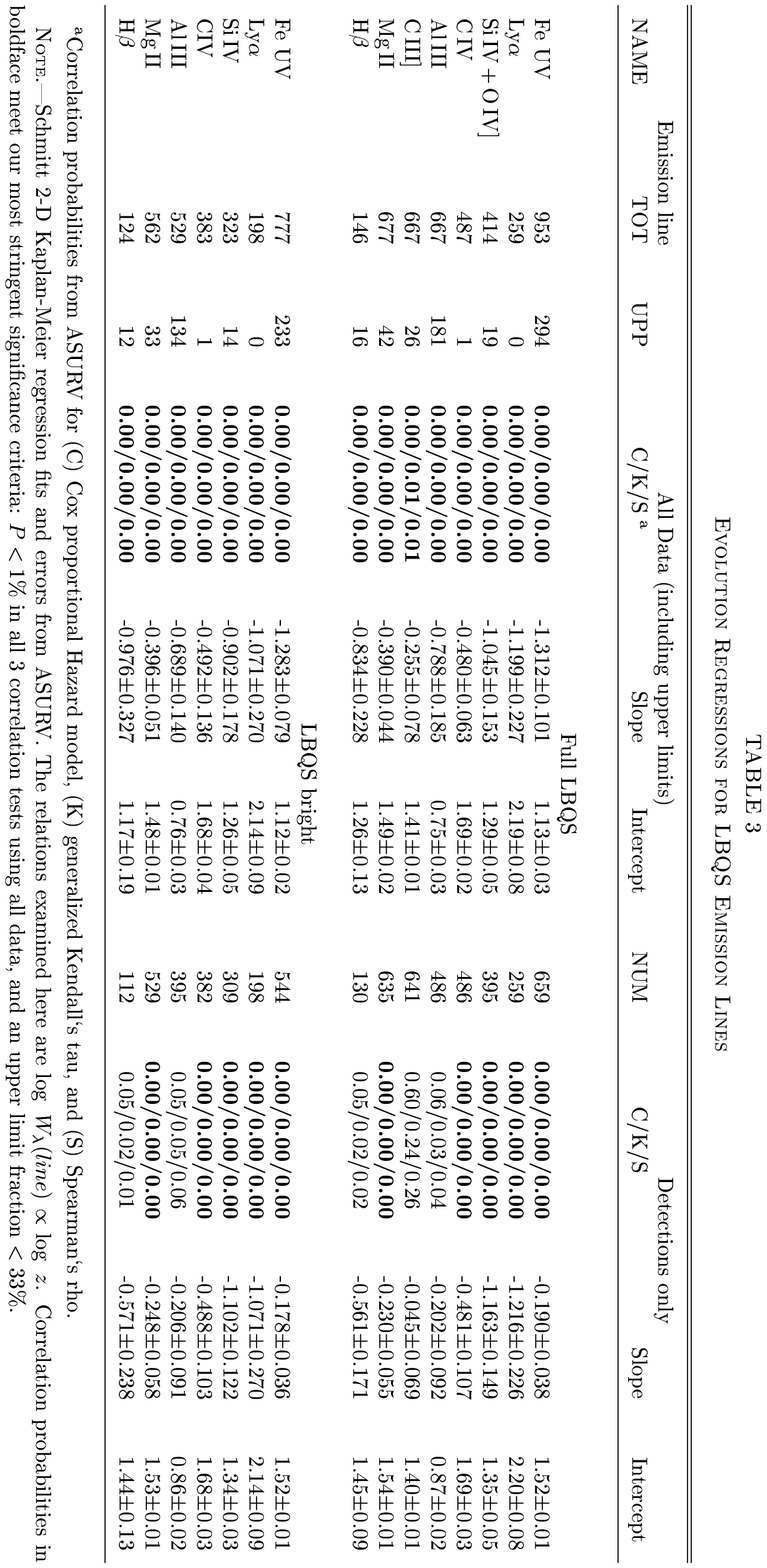}{16cm}{180}{100}{100}{300}{600}
\end{figure}

\clearpage

\begin{figure}[ht]
\figurenum{1}
\plotfiddle{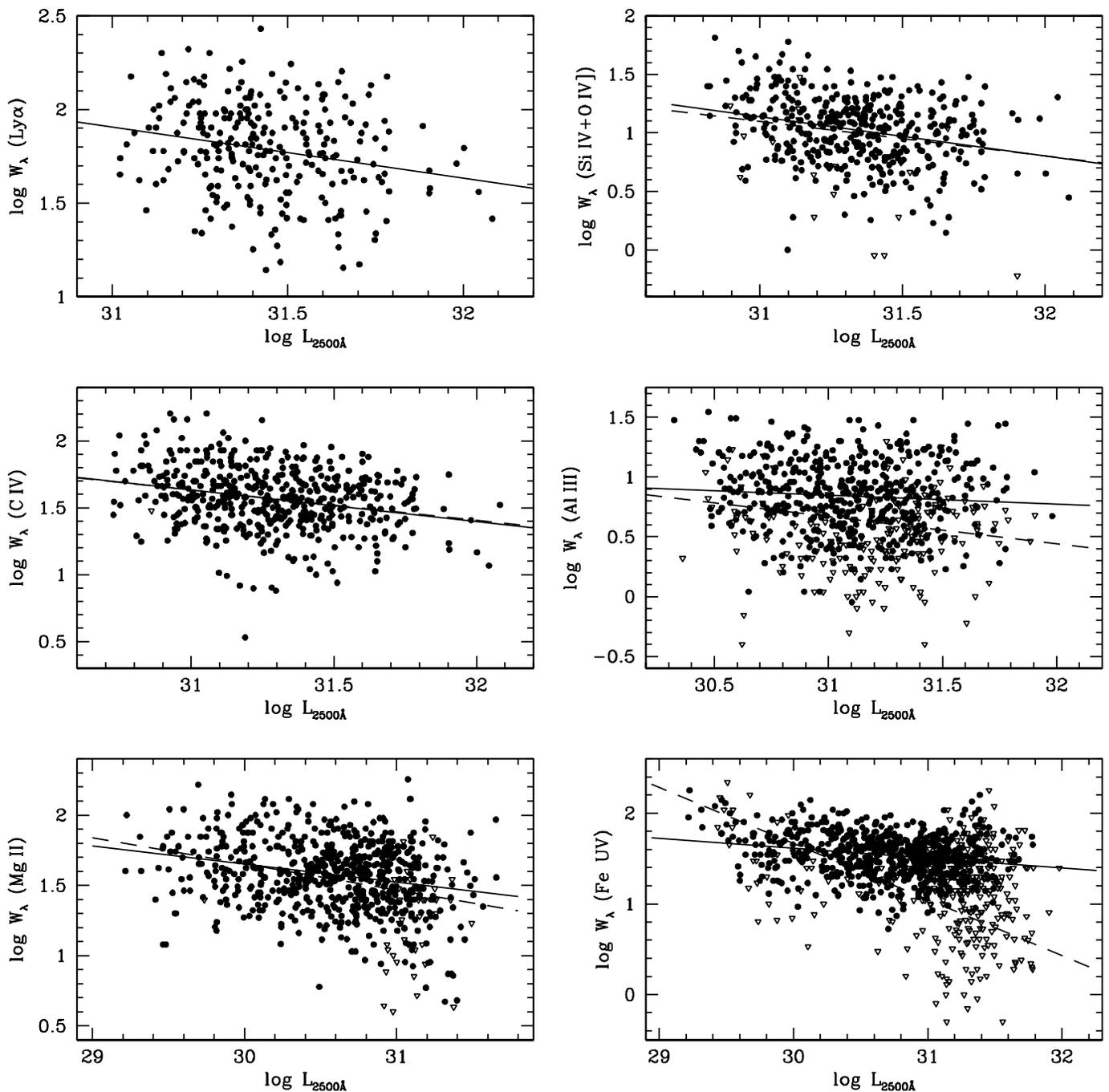}{16cm}{0}{100}{100}{-300}{-250}
\vspace{3cm}
\caption{Significant log-log correlations between the emission line
equivalent width and the monochromatic luminosity at 2500\AA\ (in
erg~s$^{-1}$~cm$^{-2}$~Hz$^{-1}$) in the LBQS sample. Filled circles
denote detections, open triangles upper limits in \ew.  Dashed
lines are best-fit linear regression including upper limits.
Solid lines show the best fit to detections only.  In cases where
there are few constraining upper limits, the two lines overlap.}
\label{fig-beff} 
\end{figure}

\clearpage
\ 
\vspace{4cm}
\begin{figure}[ht]
\figurenum{2}
\plotfiddle{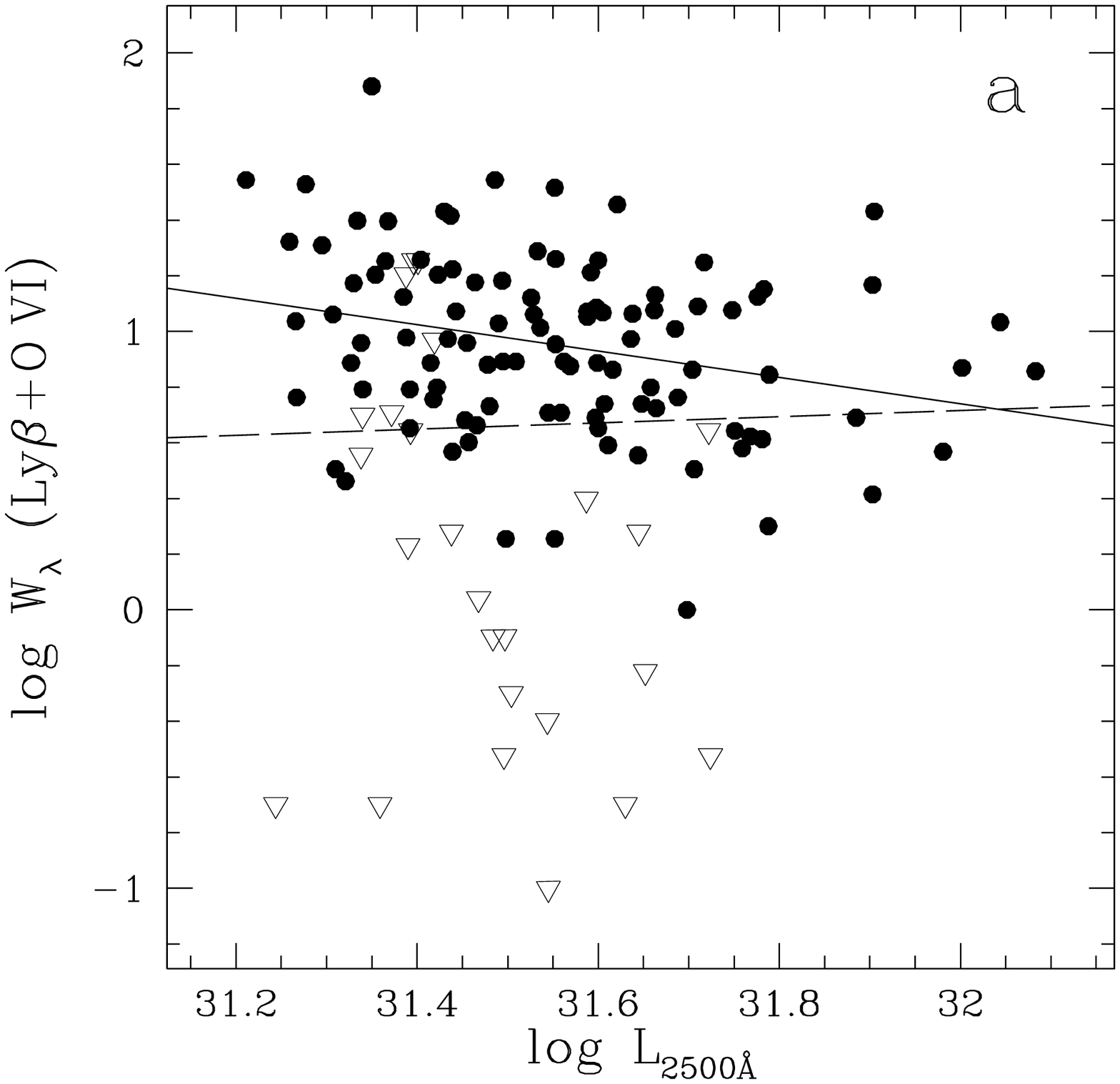}{1cm}{0}{40}{40}{-220}{-200}
\plotfiddle{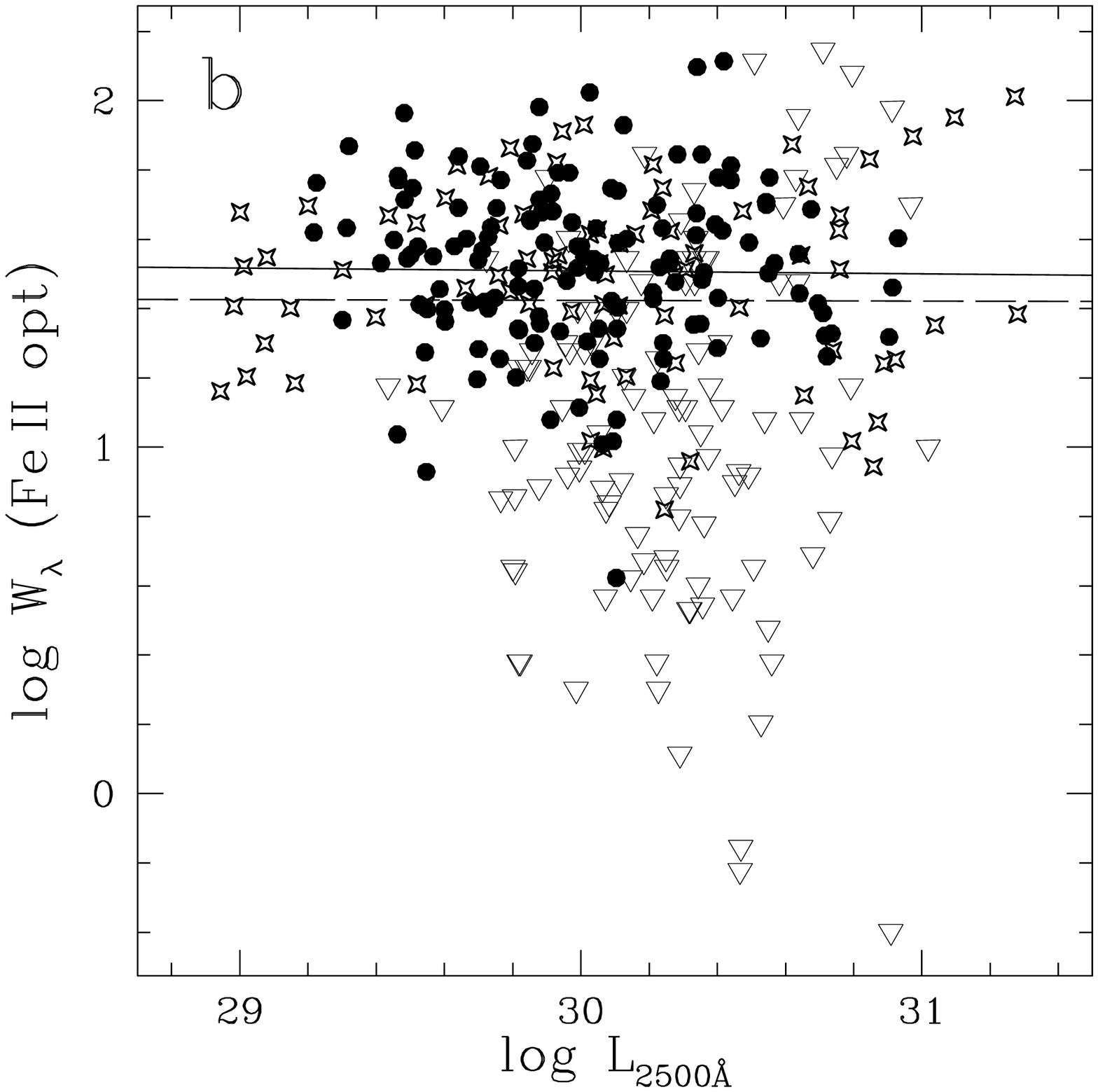}{1cm}{0}{40}{40}{   0}{-160}
\vspace{4cm}
\caption{a) The relation between the O\,VI equivalent width 
and luminosity at 2500\AA. No significant correlation is
present. However when only detections are analyzed a significant
correlation is found. The symbols have the same definition as in 
Figure~\ref{fig-beff}. b) The correlation between the \Feo\, 
equivalent width and luminosity at 2500\AA\  for the LBQS and PG
samples. Stars denote PG QSOs.}
\label{fig-beffo6}
\end{figure}
\clearpage

\begin{figure}[ht]
\figurenum{3}
\plotfiddle{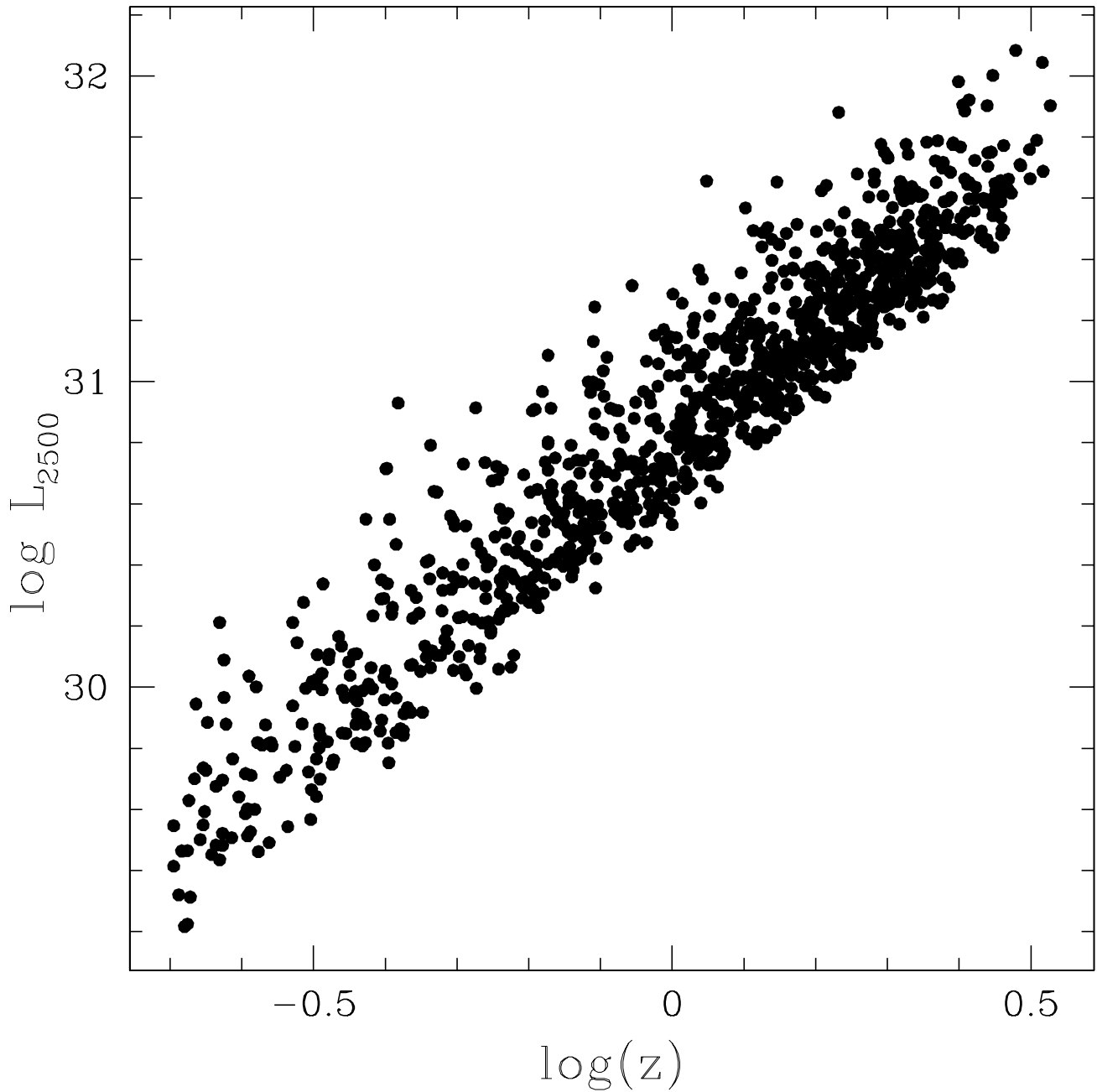}{16cm}{0}{90}{90}{-300}{-200}
\vspace{3cm}
\caption{The logarithm of the monochromatic luminosity at 2500\AA\
(in erg~s$^{-1}$~cm$^{-2}$~Hz$^{-1}$) plotted against logarithm
of the redshift $z$ for QSOs in the LBQS sample.  \loglopt\,
is derived as decribed in \S~\ref{beff}. }
\label{fig-lz} 
\end{figure}

\clearpage

\begin{figure}[ht]
\figurenum{4}
\plotfiddle{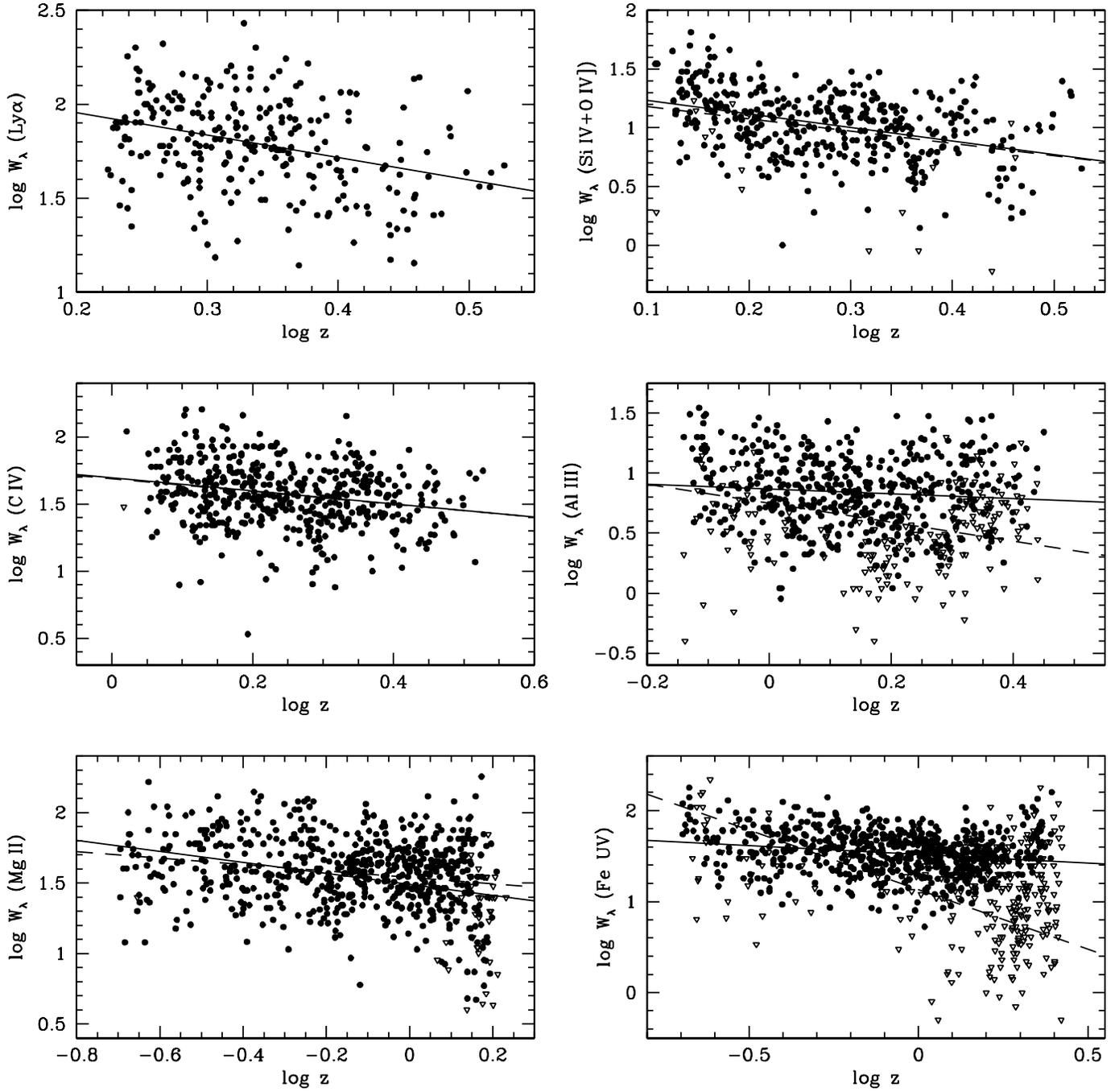}{16cm}{0}{100}{100}{-300}{-250}
\vspace{3cm}
\caption{Significant log-log correlations between the emission line
equivalent width and redshift in the LBQS sample. Symbols and
regression lines have same meaning as in Figure~\ref{fig-beff}.  The
dispersion in the panels is lower than in Figure~\ref{fig-beff}, as
reflected in the multivariate analysis results described in
\S~\ref{evol} and Table~2.}  
\label{fig-evol} 
\end{figure}

\clearpage
\begin{figure}[ht]
\figurenum{5}
\plotfiddle{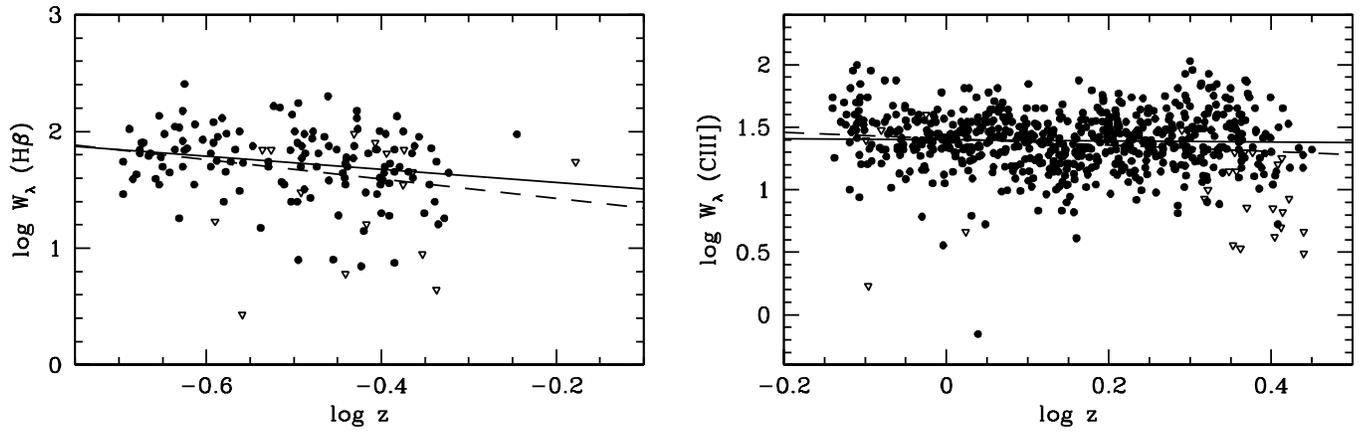}{16cm}{0}{100}{100}{-300}{-250}
\vspace{-8cm}
\caption{Significant log-log correlations between the emission line
equivalent width and redshift in the LBQS sample for
\ciii\, and \hb.  Symbols and regression lines have same meaning as in
Figure~\ref{fig-beff}.  These emission lines do not show a significant
correlation with luminosity in the LBQS sample, or with redshift
when detections alone are analyzed.}
\label{fig-evolb} 
\end{figure}

\clearpage
\begin{figure}[ht]
\figurenum{6}
\plotfiddle{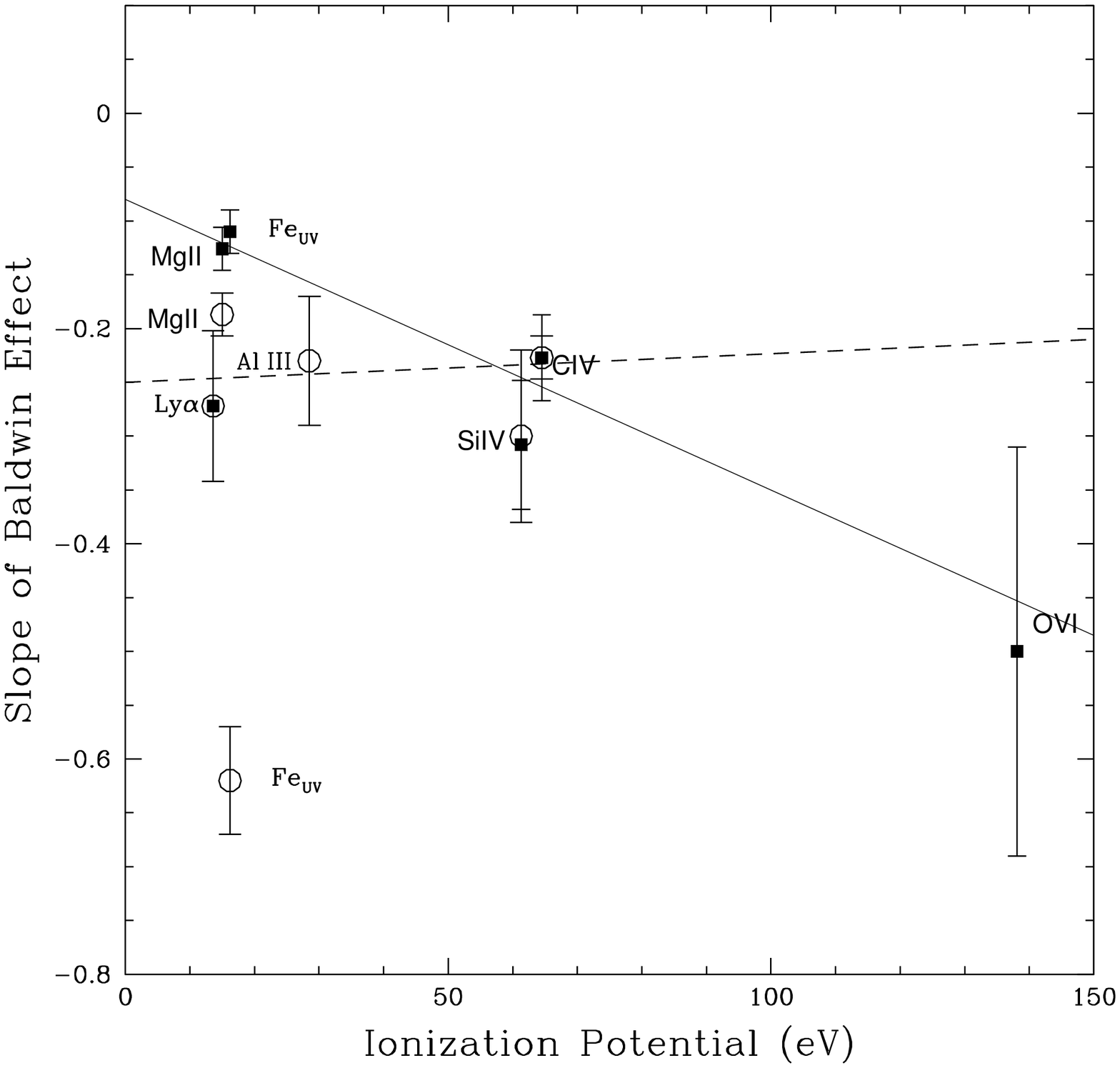}{12cm}{0}{70}{70}{-200}{-200}
\vspace{4cm}
\caption{The slope of the Baldwin Effect as a function of ionization
energy for line species with significant measured luminosity
correlations. Open symbols (fit with the dashed line) represent the
slopes of the Baldwin effect obtained from all data (including
detections and upper limits). Solid symbols (fit with the solid line)
represent the slope of the Baldwin effect for detections only.  For
the SiIV+OIV] blend, we use the mean of the ionization potentials for
both ions.  Statistically, the correlation is not significant,
but any trend is heavily dependent on measurements of O\,VI.
}
\label{fig-espey}
\end{figure}


\begin{references}
\reference{} Arav, N, Korista, K.T., de Kool, M., Junkkarinen, V. T.,
\& Begelman, M. C.  1999, ApJ, 516, 27  
\reference{} Baldwin, J.A. 1977,  ApJ, 214, 679
\reference{} Baldwin, J.A., Wampler, E. J., \& Gaskell, C. M. 1989,
ApJ, 338, 630 
\reference{} Baldwin, J. A.  1999, in {\em Proceedings of 
Quasars as Standard Candles for Cosmology}, eds. J. Baldwin
\& G. J. Ferland (San Francisco: Astronomical Society of the Pacific)
v162, 475
\reference{} Boroson, T.A., \& Green, R.F. 1992, ApJS, 80, 109 (BG92)
\reference{} Boyle, B. J.,  Shanks, T., \& Peterson, B. A. 1988, MNRAS,
243, 231 
\reference{} Brandt, W. N. 2000, NewAR, 44, 461
\reference{} Brotherton, M. S., Tran, H. D.,  Becker, R. H.,  Gregg, M. D.,
Laurent-Muehleisen, S. A., \&  White, R. L. 2001, ApJ, in press. 
\reference{} Collin, S. \& Joly, M. 2000, NewAR, 44, 531
\reference{} Dietrich, M.\& Wilhelm-Erkens, U. 2000, A\&A, 354, 17
\reference{} Djorgovski, S., Piotto, G., Capaccioli, M. 1993, AJ, 105, 2148
\reference{} Eskridge, P. B., Fabbiano, G., \&  Kim, D.-W. 1995, ApJS, 97, 141
\reference{} Espey, B. \&  Andreadis, S. 1999, in {\em Proceedings of 
Quasars as Standard Candles for Cosmology}, eds. J. Baldwin
\& G. J. Ferland (San Francisco: Astronomical Society of the Pacific)
v162, p351 (EA99)
\reference{} Fan, X., White, R. L., Davis, M., et al. 2000, AJ, 120, 1167
\reference{} Ferland, G. J. et al. 1996, ApJ, 461, 683
\reference{} Forster, K, Green, P. J., Aldcroft, T.,
Vestergaard, M., \& Foltz, C. B. 2001, ApJS, 134, in press (Paper\,I)
\reference{} Garnavich, P. M., Kirshner, R. P., Challis, et al. 1998, ApJ,
493, 53
\reference{} Giveon, U., Maoz, D., Kaspi, S., Netzer, H., \&
Smith, Paul S. 1999, MNRAS, 306, 637
\reference{} Green, P. J., et al. 1995, ApJ, 450, 51
\reference{} Green, P. J. 1996, ApJ, 467, 61
\reference{} Green, P. J. 1998, ApJ, 498, 170
\reference{} Green, P. J. 1999, in {\em Proceedings of 
Quasars as Standard Candles for Cosmology}, eds. J. Baldwin
\& G. J. Ferland (San Francisco: Astronomical Society of the Pacific)
v162, p351 
\reference{} Hamann, F. \& Ferland, G.J. 1993, ApJ, 418, 11 (HF93)
\reference{} Hamann, F. \& Ferland, G. J. 1999, ARAA, 37, 487  
\reference{} Hawkins, M. R. S. 2000, A\&AS, 143, 465
\reference{} Helfand, D. J., Stone, R.P.S., Willman, B., White, R. L.,
Becker, R. H., Price, T., Gregg, M. D., \& McMahon, R.G. 2001, AJ, in press
\reference{} Hillier, D. J. \& Miller, D. L. 1998, ApJ, 496, 407
\reference{} Iwasawa, K. \& Taniguchi, Y. 1993 , ApJ, 413, 15
\reference{} Kendall, M., \& Stuart, A. 1976, The Advanced Theory of
Statistics, Vol. II (New York: Macmillan). 
\reference{} Kinney, A. L., Rivolo, A.R., \& Koratkar, A.P. 1990, ApJ,
357, 338
\reference{} K\"oppen, J. \& Arimoto, N. 1990, A\&A, 240, 22
\reference{} Korista, K. T., Hamann, F. W., Ferguson, J. W., \&
Ferland, G. J. 1996, ApJ, 461, 641 
\reference{} Korista K. T., Baldwin J. A., \& Ferland, G. J. 1998, ApJ, 507, 24
\reference{} Krolik, J. H., \& Kallman, T. R. 1988, ApJ, 324, 714 
\reference{} Krolik J. \& Voit G. M. 1998. Ap. J. 497, L5
\reference{} Kuraszkiewicz, J., Green P. J., Forster, K., Evans, I.,
\& Koratkar, A. 2001, in preparation 
\reference{} Krolik, J. H., \& Kallman, T. R. 1988, ApJ, 324, 714 
\reference{} Kuhn, O., Elvis, M., Bechtold, J., \& Elston, R. 2001,
ApJS, submitted
\reference{} Lavalley, M., Isobe, T., \& Feigelson, E. D. 1992, ADASS, 1, 245
\reference{} Laor, A. et al. 1995, ApJS, 99, 1
\reference{} Mathur, S. 2000a, NewAR, 44, 7-9
\reference{} Mathur, S. 2000b, MNRAS, 314, 17
\reference{} Mathur, S., Matt, G., Green, P. J., Elvis, M., \& Singh,
K. P. et al. 2001, in preparation.
\reference{} Melchiorri, A., Ade, P. A. R., de Bernardis, P.,  et
al. 2000, ApJL, 536, 63 
\reference{} Morris, P., Conti, P. S., Lamers, H. J. G. L. M.,
Koenigsberger, G. 1993, ApJL, 414, 25
\reference{} Murray, N. \& Chiang, J 1998, ApJ, 494, 125
\reference{} Netzer, H., Laor, A., \& Gondhalekar, P.M. 1992, MNRAS
254, 15
\reference{} Osmer, P.S., Porter, A.C., \& Green, R.F. 1994, ApJ, 436, 678
\reference{} Park, C., Colley, W. N., Gott, J., R., Ratra, B.,
Spergel, D. N., \& Sugiyama, N. 1998, ApJ, 506, 473
\reference{} Perlmutter, S. et al. 1999, ApJ, 517, 565
\reference{} Pogge, R.W., \& Peterson, B. 1992, AJ, 103, 1084
\reference{} Reeves, J. N. et al. 2001, A\&A 365, L116
\reference{} Richards, G. T. et al. 2001, AJ, in press
\reference{} Sergeev, S. G., Pronik, V. I., Sergeeva, E. A., Malkov,
Y. F. 1999, AJ, 118, 2658
\reference{} Shields, J. C., Ferland, G. J., \& Peterson. B. M.
1995, ApJ, 441, 507
\reference{} Thompson, K. L., Hill, G. J., Elston, R. 1999, ApJ, 515,
487 
\reference{} Tytler, D., \& Fan X.-M. 1992, ApJS, 79, 1
\reference{} Vestergaard, M., \& Wilkes, B.J. 2000, ApJ, submitted.
\reference{} Vernet, J., Fosbury, R. A. E., Villar-Martin, M., Cohen,
M. H., Cimatti, A., di~Serego~alighieri, S., \& Goodrich, R. W. 2001,
A\&A, 366, 7
\reference{} Villar-Martin, M., Vernet, J., Fosbury, R. A. E., 
Binette, L., Tadhunter, C. N., \& Rocca-Volmerange, B. 1999,
A\&A, 351, 47
\reference{} Webb, W. \& Malkan, M. 2000, ApJ, 540, 652
\reference{} Wheeler J. C., Sneden C., \& Truran J. W. 1989, ARAA, 27, 279  
\reference{} Wandel, A. 1999a, ApJ, 527, 649
\reference{} Wandel, A. 1999b, ApJ, 527, 657
\reference{} Wandel, A. 1999c, ApJL, 519, 39
\reference{} Wilkes, B. J., Kuraszkiewicz, J., Green, P. J.,
Mathur, S., \& J.C. McDowell 1999,   ApJ, 513, 76
\reference{} Wilkes, B. J. et al. 2001, in {\em New Era of Wide Field
Astronomy}, eds. Clowes,  R.G.,  Adamson, A.J., \& Bromage, G.E. (San
Francisco: Astronomical Society of the Pacific), in press 
\reference{} Wilkes, B. J., Tananbaum, H., Worrall, D. M., Avni, Y.,
Oey, M. S., \& Flanagan, J. 1994, ApJS, 92, 53 
\reference{} Wills, B.J., Netzer, H., \& Wills, D. 1985, ApJ, 288, 94
\reference{} Wills, B.J., Laor, A., Brotherton, M. S., Wills, D.,
Wilkes, B. J., Ferland, G. J., \& Shanf, Z. 1999, ApJ, 515, L53
\reference{} Wilman, R. J. \& Fabian, A. C. 1999, ApJ, 522, 157
\reference{} Wu, C.-C., Boggess, A., \& Gull, T. R 1983, ApJ, 266, 28
\reference{} York, D., et al. 2000, AJ, 120, 1579
\reference{} Zamorani, G., Marano, B., Mignoli, M., Zitelli, V., \&
Boyle, B. J. 1992, MNRAS, 256, 238
\reference{} Zheng, W., Fang, L.-Z., \& Binette, L. 1992, ApJ, 392, 74
\reference{} Zheng, W., Kriss, G.A., \& Davidsen, A.F. 1995, ApJ, 440,
606


\end{references}
\end{document}